\documentclass[aip,preprint]{revtex4}
\usepackage{graphicx}
\usepackage{epsfig}
\usepackage{bm}
\usepackage{amsmath}
\usepackage{amssymb}
\usepackage{wasysym}
\usepackage{epstopdf}
\usepackage{color}
\usepackage{xcolor}
\usepackage{epstopdf}
\usepackage{upgreek}
\usepackage{color}

\usepackage[colorlinks=false, bookmarks=true]{hyperref}

\newcommand{\etal}{\emph{et al.}}

\begin{document}


\title[Microdroplet impact at very high velocity]{Microdroplet impact at very high velocity} 
\author{Claas Willem Visser}
\email{c.visser@utwente.nl}
\author{Yoshiyuki Tagawa}
\author{Chao Sun}
\email{c.sun@utwente.nl}
\author{Detlef Lohse}
\email{d.lohse@utwente.nl}

\affiliation{ 
Physics of Fluids Group, Faculty of Science and Technology, J. M. Burgers Centre for Fluid Dynamics, University of Twente, The Netherlands}

\date{\today}

\begin{abstract}
Water microdroplet impact at velocities up to 100 m/s for droplet diameters from 12 to 100 $\upmu$m is studied. This parameter range covers the transition from capillary-limited to viscosity-limited spreading of the impacting droplet. 
Splashing is absent for all measurements; the droplets always gently spread over the surface. 
The maximum spreading radius is compared to several existing models. 
The model by Pasandideh-Fard \etal~\cite{pasandideh96} agrees well with the measured data, indicating the importance of a thin boundary layer just above the surface, in which most of the viscous dissipation in the spreading droplet takes place. As explained by the initial air layer under the impacting droplet, a contact angle of 180 degrees is used as model input. 
\end{abstract}

\maketitle

\section{Introduction}

High-pressure spray cleaning, droplet-wall interactions in diesel engines, and plasma spraying are notable examples of processes in which high-speed impact of small droplets on a solid surface is a key phenomenon. In these applications, droplets with a characteristic size of 1 to 100 $\upmu$m and a velocity of order 100 m/s impact on a solid surface \cite{mitra01,blaisot05,haller02}. Despite this industrial interest, microscale droplet impact at very high velocity ($U_0>50$ m/s) has only been studied for solidifying metal droplets \cite{mcdonald06}. This is mainly due to the challenging parameter regime: very high spatial and temporal resolutions are required to study the relevant phenomena. In addition, it is difficult to create impact events at these velocities.
An understanding of the phenomena of high-speed micro-sized droplet impact is thus lacking \cite{rein93,yarin06}.

In this work, we aim to extend current results in three ways. First, we present a novel high-velocity droplet generation, by using a method to create ultrafast liquid jets. Second, using high-speed imaging, the impact dynamics is studied. Third, a quantitative investigation of the maximum spreading radius will be presented and compared to existing models, to improve our understanding of droplet spreading. \\

\section{Parameter space}

\begin{figure}
\includegraphics[width=0.95\textwidth]{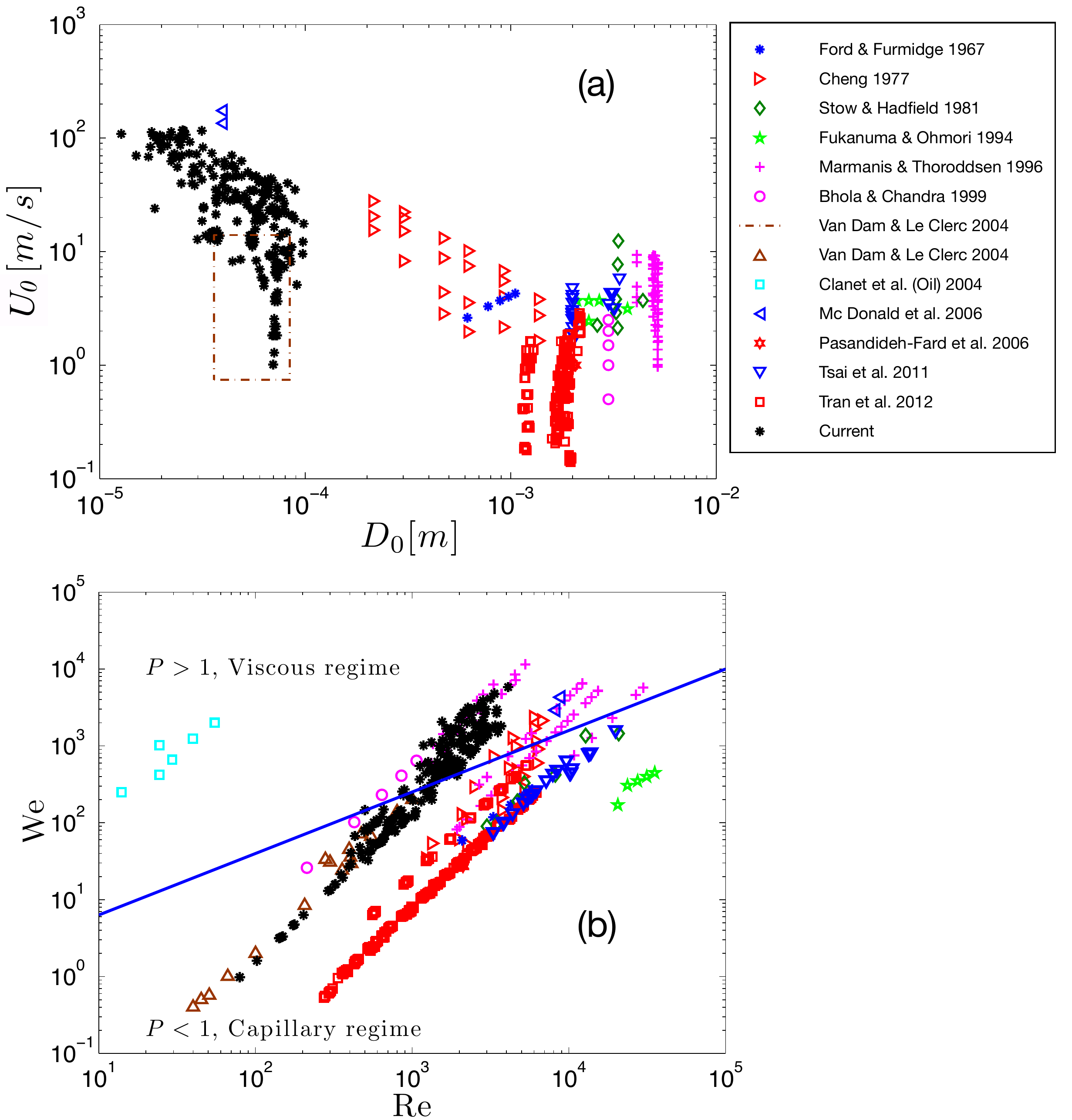}
\caption{Parameter space of (a) droplet velocities and radii and (b) Weber and Reynolds numbers  for various droplet impact experiments \cite{bhola99,cheng77,clanet04,ford67,fukanuma94,marmanis96,mcdonald06,pasandideh96,stow81,dam04,tsai11,tran11}.  
The solid blue line signals $P$ = 1 and separates the capillary regime with $P <$ 1 from the viscous regime with $P >$ 1.}
\label{fig:parameterspace}
\end{figure}

To compare our results to previous work, a phase diagram of the droplet size ($D_0$) and impact speed ($U_0$) is plotted in figure~\ref{fig:parameterspace} (a). 
Most work up to now has focused on impact dynamics of droplets with a size of $\sim$1 mm in diameter. The studies on microdroplet impact were mainly at relatively low speed (up to 10 m/s)\cite{dam04,mcdonald06,cheng77}. Our study connects these previous investigations, in particular those of water microdroplets at lower velocies \cite{dam04} and metal microdroplet impact at very high velocities \cite{mcdonald06}. 

Figure~\ref{fig:parameterspace} (b) shows a phase diagram of the achieved Reynolds and Weber numbers of experimental droplet studies. 
The Weber number is defined as $\mathrm{We} = \rho D_0 U_0^2/\sigma$, where $\sigma$ the surface tension and $\rho$ the density. The Reynolds number is given by $\mathrm{Re} = \rho D_0 U_0/\mu$ where $\mu$ represents dynamic viscosity.
As shown in figure~\ref{fig:parameterspace} (b), the Re-We values of our measurements largely overlap with previous data. However, our data have been taken for microdroplets instead of mm-sized droplets as indicated in figure~\ref{fig:parameterspace} (a). 
As the droplet size is a key control parameter for impact dynamics, various models developed to describe the mm-sized droplet impact may not hold for microdroplet impact dynamics.
The large impact velocity also explains why our data have relatively high Weber numbers for given Reynolds numbers.

Our data cover the transition between the capillary- to the viscosity-dominated limits of droplet spreading \cite{clanet04}, as shown by the solid line in figure \ref{fig:parameterspace} (a) (an explanation of this transition is provided below). So far, to investigate this transition, liquids of different viscosities and surface tension were required in order to achieve a sufficient coverage of the Re-We parameter space. Using droplets of microscopic scales lowers the Weber number for which this transition takes place, allowing to study the transition region with a single liquid. 

\section{Experimental setup}
\begin{figure}
\label{fig:dropGeneration}
\centering
\includegraphics[width=0.7\textwidth]{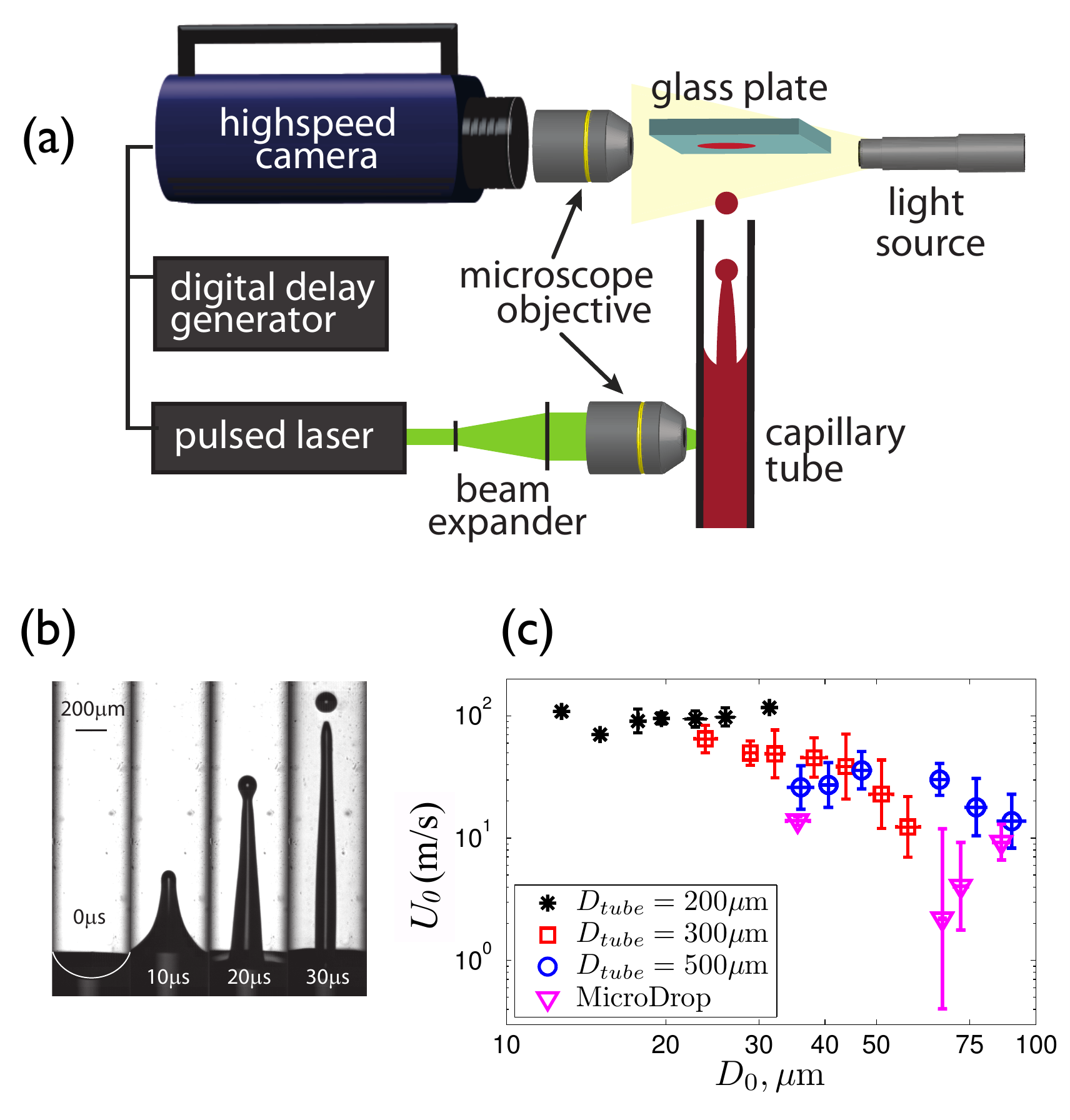}
\caption{(a) Setup used for the generation of fast droplets. (b) Jet generation and breakup at different instants. (c) Range of the achieved droplet velocities and radii, for different droplet generator settings. Capillary tube diameters are decreased from 500 $\mu$m to 200 $\mu$m to generate smaller and faster droplets. In addition, to create slow droplets, a MicroDrop apparatus is used. As approximately 230 measurements were performed, the data are binned. The bars represent one standard deviation.}
\label{fig:setup}
\end{figure}

To create high-velocity microdroplets ($U_0\geq10$m/s), we make use of a new method to create ultrafast liquid jets \cite{tagawa12}, as sketched in figure \ref{fig:setup} (a). In a nutshell, the method works as follows. By focusing a laser pulse with a microscope objective, a vapor bubble is created in a capillary tube, by laser-induced cavitation. From this bubble, a shock wave travels to the meniscus, which in turn forms a liquid jet thanks to flow focusing. Subsequently, this jet breaks up into tiny droplets with a velocity similar to the jet velocity (figure \ref{fig:setup} (b)). 

It is found that the tube diameter is a key control parameter for both the jet diameter and the jet tip velocity \cite{tagawa12}. Therefore, capillaries with diameters are used to create a range of droplet sizes and velocities. In addition, the laser energy and the distance between the laser focus and the meniscus are varied to generate droplets at different velocities for a given tube diameter \cite{tagawa12}, resulting in the diameter- and velocity ranges, as shown in figure \ref{fig:setup} (c). 
With this method, approximately 170 droplet impacts on a dry surface were examined. For the sake of clarity, the data is binned in the figure.

To create droplets at velocities between 1 and 10 m/s, a commercially available Microdrop dispenser is used. By varying the input voltage between 60 V and 160 V per pulse, a range of velocities is covered. The droplet diameter ranges from 40 $\upmu$m to 80 $\upmu$m, as shown in figure  \ref{fig:setup} (c).

A standard microscope slide was used as an impact plate, which is placed above the tip of the capillary tube. Atomic force microscope measurements indicated a roughness $R_a$ below 10 nm. The droplet impact is visualized from the side. After each impact measurement, the glass plate was cleaned with ethanol and dried with paper tissue. Frequent checks with a 10$\times$ optical microscope (after the cleaning procedure) indicated that this method usually resulted in an optically clean surface, i.e., hardly any paper fibers or dirt were sticking to the surface. 
The liquid used in the present work was a standard water-based blue inkjet printer ink, with density $\rho = 998$ kg/m$^3$, surface tension $\sigma = 72$ mN/m$^2$, and viscosity $\mu = 10^{3}$ Pa$\cdot$s.

Magnifications of 10$\times$ to 40$\times$ are obtained by combining a standard Olympus 10$\times$ objective with an adjustable $12\times$ zoom lens (Navitar 1-50015). An Olympus ILP-1 light source is used for illumination. 
An ultra high-speed camera (Shimadzu HPV-1) is used to study the microdroplet impact dynamics, at recording rates of $1.25\cdot10^5$ to $10^6$ frames per second. 
For the fastest droplets ($U_0 \approx 100$ m/s), the impact duration is approximately $\tau = D_0/U_0 \approx$ 0.2 $\upmu$s, which is below the temporal resolution of the camera. Thus, completely capturing of such events requires even higher frame rates \cite{chin03} or pulsed illumination, e.g. as used in Ref. \cite{bos11}. However, these techniques require a level of control of the moment of impact which is not achieved with the present setup. The high-speed camera has a minimum shutter time of 500 ns. Therefore, at high impact velocities, substantial motion blur is observed (e.g. figure \ref{fig:impact-outcome} (c)), as the shutter time approaches $\tau$. Still, with the current setup, the presence of splashing could be assessed and the maximum spreading radius could be observed in an entirely new parameter regime. \\

\section{Results}

\subsection{Impact Phenomenology} 

\begin{figure*}
(a)\includegraphics[width=0.55\textwidth]{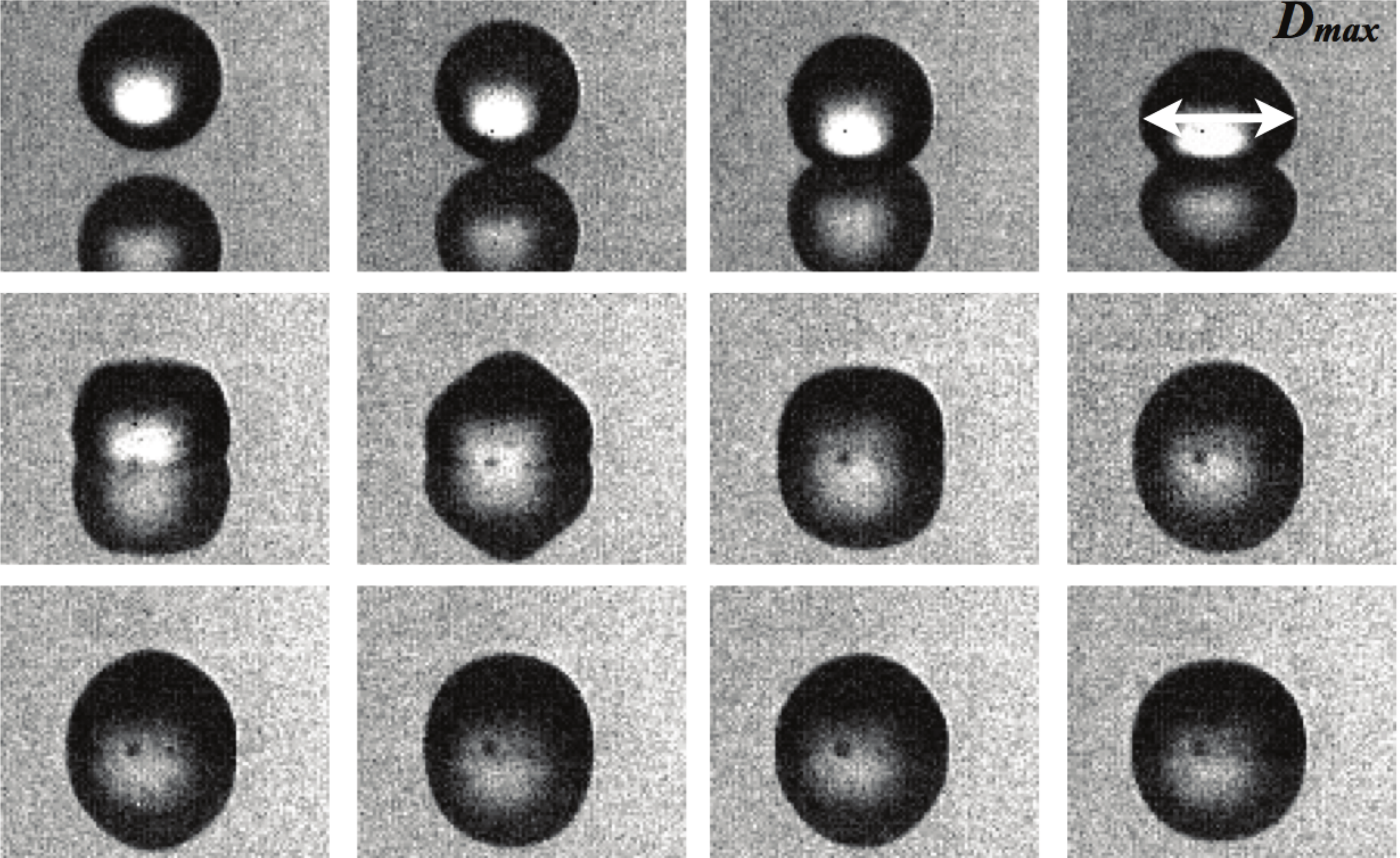}
(b)\includegraphics[width=0.65\textwidth]{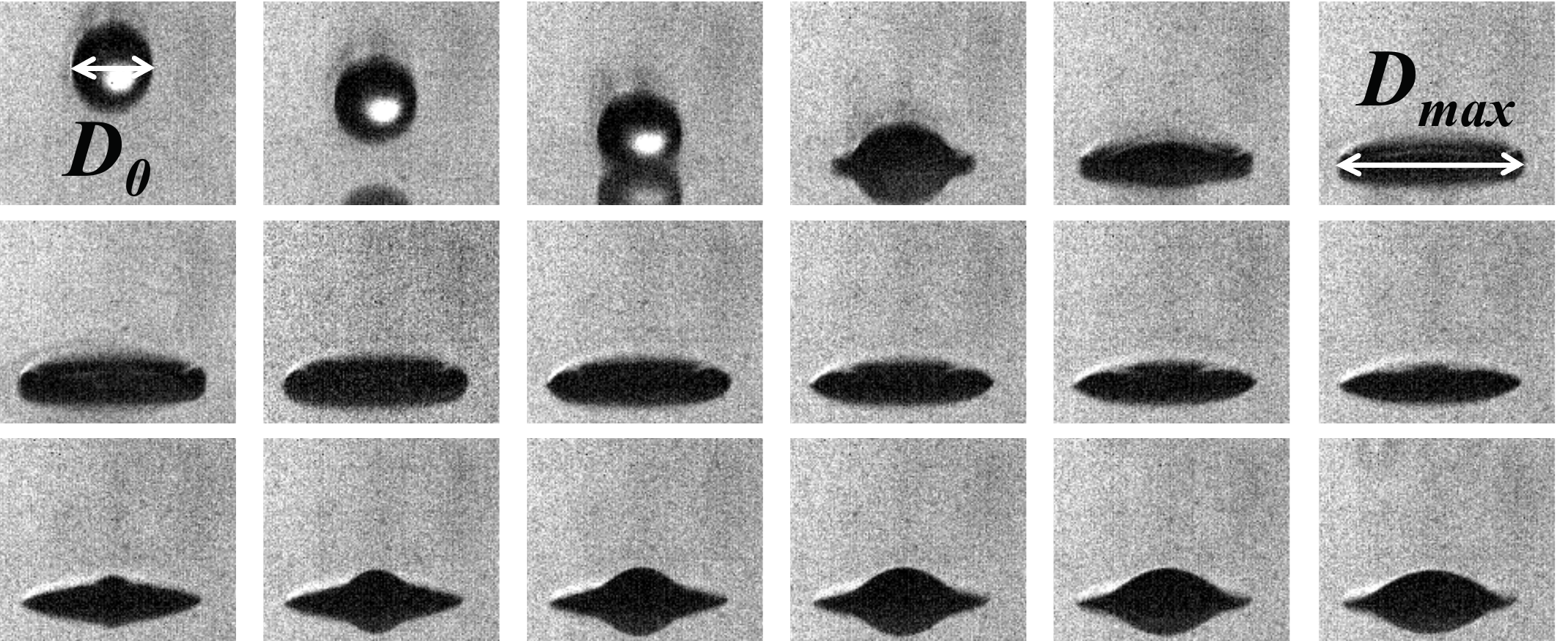}
(c)\includegraphics[width=0.44\textwidth]{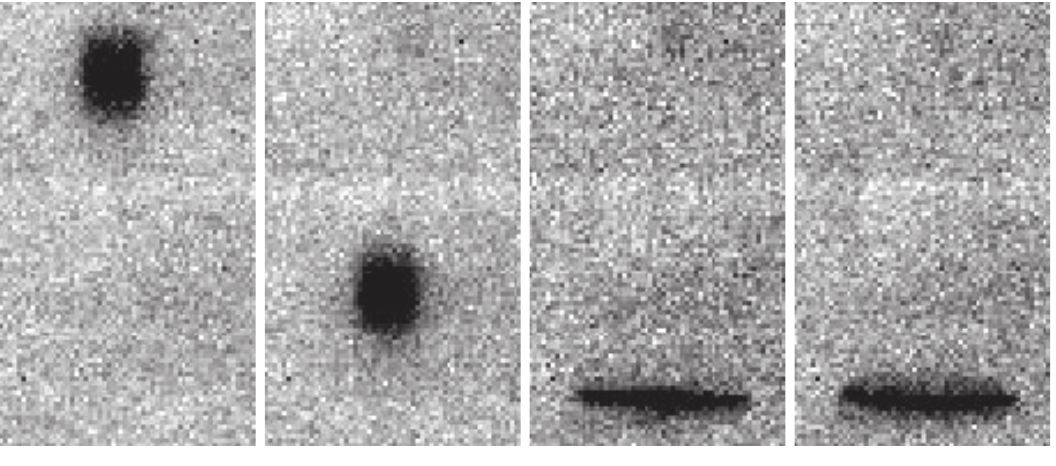}
(d)\includegraphics[width=0.45\textwidth]{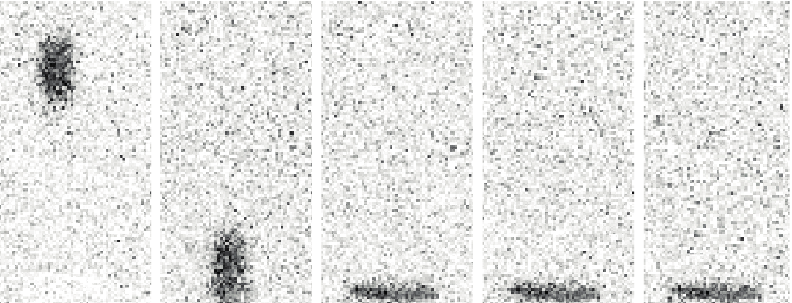}
\caption{Time series of droplet impact at different impact velocities. (a) $U_0=0.7$m/s, $D_0=70\upmu$m, $\mathrm{We} = 0.52$, $\mathrm{Re} = 56$, and the time lag $\delta t$ between the frames is 8 $\upmu$s. Droplet oscillations and air bubble entrapment can clearly be seen. 
(b) $U_0=7.7$m/s, $D_0=71\upmu$m, $\mathrm{We} = 60$, $\mathrm{Re} = 613$, and $\delta t=4\upmu$s. Spreading into a thin center sheet and a thicker rim are observed, followed by oscillations and partial withdrawal of fluid from the rim towards the droplet center. 
(c) $U_0=73$m/s, $D_0=23\upmu$m, $\mathrm{We} = 1.8\cdot10^3$, $\mathrm{Re} = 1.9\cdot10^3$,  and $\delta t=1\upmu$s. The details of the spreading phase can no longer be resolved. 
(d) $U_0=100$m/s, $D_0=20\upmu$m, $\mathrm{We} = 3\cdot10^3$, $\mathrm{Re} = 2.3\cdot10^3$, and $\delta t=1\upmu$s.
}
\label{fig:impact-outcome}
\end{figure*}

In general, the droplet impact process can be divided into the following phases. When the droplet approaches the solid surface, the air between the falling drop and the surface is strongly squeezed, leading to a pressure buildup in the air under the drop. The enhanced pressure results in a dimple formation in the droplet and an air layer development between the droplet and the target plate \cite{man10,vee12,bou12}. Before the droplet wets the surface, the liquid moves on top of this air cushion. The droplet extends in the radial direction until it reaches a maximum spreading radius. In this phase, splashing can occur \cite{xu05,yarin06,man10,dri11,tsai11,man12}. 
Finally, the droplet completely wets the surface and an air bubble is entrapped \cite{bou12}.

The first aim of this work is to assess the dynamics of microdroplet impact, i.e. to find out whether the drop is in the splashing or gentle spreading regime. The latter one is defined by droplet deformation into a pancake shape, without satellite droplet formation. Time series of droplet impacts and the subsequent spreading phases are shown in figure \ref{fig:impact-outcome}. For these figures, the velocity ranges from 0.7 m/s to 100 m/s.

At low velocity ($U_0 = 0.7$ m/s), initial flattening of the droplet bottom is observed, as shown in figure \ref{fig:impact-outcome} (a). Subsequently, the droplet spreads over the substrate into a (virtually) half-dome shaped cap. During spreading, the droplet starts to oscillate and comes to rest only after the droplet has reached its maximum diameter. Additionally, air bubble entrainment is observed (the small black dot, just left of the droplet center). These phenomena are consistent with what had been reported in Ref. \cite{dam04}, which includes a detailed discussion on the droplet oscillation frequency and the size and cause of the bubble/cushioning entrained \cite{tho05,tsai11,vee12,bou12}. 
At medium velocities (7.7 m/s, figure \ref{fig:impact-outcome} (b)), the droplet deforms into a disc-like structure. Here, the central impact area is a sheet-like structure surrounded by a thicker rim. Again, capillary oscillations were observed.  
Finally, even at very high velocities (figures \ref{fig:impact-outcome} (c) and \ref{fig:impact-outcome} (d)), still no splashing is observed. Thus, we conclude that gentle impact occurs for all velocities and droplet sizes investigated in the present work. 

\subsection{Maximum spreading} 
Now the maximum spreading radius will be determined and compared against various models.
As knowledge of the maximum spreading diameter is of paramount importance for industrial applications, a plethora of models has been developed \cite{cheng77,clanet04,pasandideh96,eggers10,chandra91,bechtel81,madejski76,roisman02}. It is generally agreed on that the spreading is limited by either viscosity or surface tension. Therefore, a key issue is to define which of these is dominant. To study this issue, 
we will first briefly summarize several models. 

\begin{figure}
\label{fig:modelEval}
\centering
\includegraphics[width=0.5\textwidth]{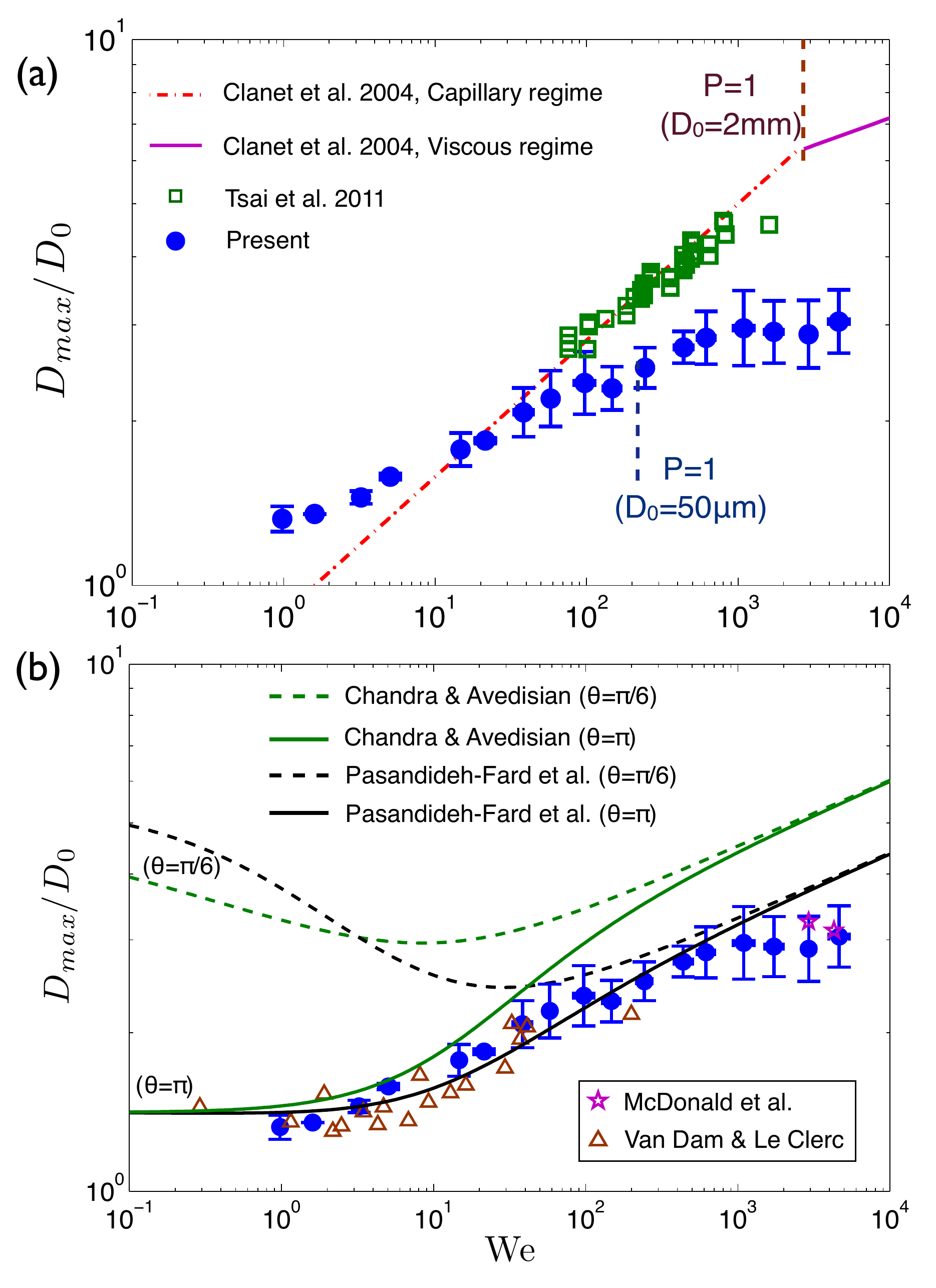}
\caption{The normalized maximum spreading $D_{max}/D_0$ vs. $\mathrm{We}$. (a) Solid circles: the present microdroplet data. Open squares: mm-sized droplets impact on superhydrophobic surfaces from Tsai {\it et al.} \cite{tsai11} . Dash-dot line: the capillary model by Clanet {\it et al.} \cite{clanet04}. Solid line: the viscous model by Clanet {\it et al.} \cite{clanet04}. The short vertical  dashed lines show the transitional Weber number $\mathrm{We_t}$ at $P$ = 1 for a 2 mm-sized droplet and for a 50 $\upmu$m droplet. 
(b) Solid circles: the present microdroplet data. Stars: data from Mc Donald {\it et al.} \cite{mcdonald06} with impact of molten metal microdroplets. Open triangles: low-speed microdroplet impact experiments by Van Dam \& Le Clerc \cite{dam04}.
Lines: Model by Chandra \& Avedisian \cite{chandra91} (dark-green lines) and by Pasandideh-Fard \cite{pasandideh96} model (black lines), evaluated for the initial diameter $D_0 = 50 \upmu$m and contact angles $\pi/6$ (dashed lines) and $\pi$ (solid lines)}. 
\label{fig:dmax}
\end{figure}

Assuming an inviscid liquid, the maximum spreading is limited by the surface tension. Balancing the Laplace pressure force with the inertial deceleration of the drop, a scaling of $D_{max}/D_0 \sim \mathrm{We}^{1/4}$ is obtained \cite{cheng77,clanet04}. This scaling is remarkably robust \cite{clanet04,tsai11,marmanis96} in the capillary regime.
As shown in figure \ref{fig:dmax} (a), the data of mm-sized droplets impact on superhydrophobic surfaces obtained by Tsai {\it et al.} \cite{tsai11} well agree with this 1/4 scaling law.

Another limiting case is to completely ignore surface tension and assume that the maximum spreading radius is limited by the viscous dissipation during droplet spreading \cite{clanet04}. This yields $D_{max}/D_0 \sim \mathrm{Re}^{1/5}$ as scaling law, which holds well for mm-sized droplets in the viscous regime as shown in Ref. \cite{clanet04}.

To quantify the transition between the viscous and the capillary regimes, Clanet {\it et al.} \cite{clanet04} defined the parameter $P = \mathrm{We}/\mathrm{Re}^{4/5}$. For $P<1$, a surface-tension dominated regime is expected (i.e. $D_{max}/D_0 \sim \mathrm{We}^{1/4}$), whereas a viscous scaling is predicted for $P>1$, yielding the previously mentioned $D_{max}/D_0 \sim \mathrm{Re}^{1/5}$. In figure \ref{fig:dmax} (a), the transitional Weber number $We_t$, defined by $P$ = 1, for droplets of $D_0$ = 2 mm and 50 $\upmu$m are plotted. For 2 mm-sized droplets, $\mathrm{We_t} \simeq 3\times10^3$, whereas it decreases to $\mathrm{We_t} \simeq 2\times10^2$ for 50 $\upmu$m-sized ones. As shown in figure \ref{fig:dmax} (a), the present microdroplet data are in between the transition regime from the capillary regime to the viscous regime. 
For a given We, the microdroplet spreading is lower than that of mm-sized droplets due to the viscous effects.
This combination of two simple models provides a decent first description of the impact dynamics. 
However, more detailed models are possible and have been developed. 
As described by Chandra \& Avedisian \cite{chandra91}, the dissipated energy equals the work done, and can thus be estimated by 
\begin{align}
W = \int_0^\tau \int_{V_{\nu}}\phi \approx V_\nu \tau \phi \approx V_\nu \tau \mu \left(\frac{U_0}{L}\right)^2
\label{eq:1}
\end{align}
where $\phi$ the dissipation function, estimated as $\mu (U_0/L)^2$ \cite{chandra91,clanet04}, with $\tau$ the typical impact timescale, $V_\nu = \pi LD_{max}^2$ the total droplet volume, and $L$ the characteristic dissipation length scale, which in this model is selected as the height of the droplet splat $h$. In the model by Chandra \& Avedisian \cite{chandra91}, $W$ is used in an energy balance $E_k + E_s = W + E'_s$, where $E_k$ is kinetic energy, and $E_s$ and $E'_s$ are surface energies before and after impact, respectively. However, as shown by Pasanideh-Fard {\it et al.} \cite{pasandideh96} and our results, this model strongly overpredicts the maximum spreading radius.

The model was revisited by Pasandideh-Fard {\it et al.} \cite{pasandideh96}. Their numerical simulations suggested that the dissipation is taking place in a thin boundary layer within the expanding droplet, implying that the characteristic length scale
$L$ in the above eq.~\ref{eq:1} is not the pancake thickness $h$ but has to be replaced by the Prandtl-Blasius boundary layer thickness $\delta$ \cite{pasandideh96}: 
\begin{align}
\delta= 2\frac{D_0}{\sqrt{\mathrm{Re}}},
\end{align}
finally resulting in the following equation for $D_{max}/D_0$:
\begin{align}
D_{max}/D_0 = \sqrt{\frac{\mathrm{We}+12}{3(1-\cos\theta) + 4(\mathrm{We}/\sqrt{\mathrm{Re}})}}.
\end{align}
Here, $\theta$ is the contact angle. 
At low impact velocities ($\mathrm{We}\lesssim10$) this model saturates, as the droplet impact can be considered to be effectively static and $E_k$ and $W$ vanish. 

In figure \ref{fig:dmax} (b), the measured maximum spreading for the microdroplet impacts is plotted versus the Weber number. The maximum spreading is measured at the moment when the deformation of the droplet is maximum before it wets the surfaces, as shown in figure~\ref{fig:impact-outcome} (a,b). 
At low velocity, the present data saturates around a spreading of $\sim$1.3 times the initial diameter, which has a good overlap with previous low-speed microdroplet impact experiments \cite{dam04}.    
At high velocities, spreading of only $\sim$3 times the initial diameter is found for even the fastest droplets. This is consistent with (even faster) impact of molten metal microdroplets \cite{mcdonald06}, even though the liquid properties were very different from our experiments.

Figure \ref{fig:dmax} (b) is complemented with the models of Chandra \& Avedisian \cite{chandra91} and Pasandideh-Fard \cite{pasandideh96}, evaluated for an initial diameter of $D_0 = 50 \upmu$m. 
We first examine the maximum spreading calculated from these models for two different contact angles, i.e. $\theta$ = $\pi/6$ and $\pi$. The dashed lines in figure \ref{fig:dmax} (b) clearly show that the contact angle of $\pi/6$ results in a much larger spreading factor for both models as compared to the experimental data.
As discussed above, an air layer is present between the spreading drop and the solid surface before the droplet wets the surface \cite{xu05,man10,vee12,bou12}, implying a ``contact angle'' which would be best described by $\theta = \pi$.
As shown with the solid lines in figure~ \ref{fig:dmax} (b), this approach remarkably decreases the deviations between the experiments and the results of both models.  
A good agreement is found between the microdroplet data and the Pasandideh-Fard model \cite{pasandideh96} up to $\mathrm{We} \approx 10^3$. This indicates the importance of a finite boundary layer thickness in the dissipation of spreading droplets in the present parameter regime. 
For $\mathrm{We} > 10^3$, the increasing trend of $D_{max}/D_0$ versus We seems to saturate. This finding is consistent with Ref.~\cite{cheng77}, and will be studied in future work.

\section{Conclusions}
The impact of water microdroplets on a smooth solid surfaces is investigated experimentally. By using a new droplet-generating device, impact events were created at velocities from 1 to 100 m/s and droplet diameters between 12 and $100 \upmu$m. This parameter regime covers the transition between surface tension- and viscosity-dominated spreading of the droplet. 
For all impact events, no splashing is observed. The maximum spreading radius was compared to several models. The model by Pasandideh-Fard {\it et al.} \cite{pasandideh96} performs best, indicating that boundary layer dynamics play a key role in droplet spreading. In addition, we find that an initial contact angle of 180 degrees should be used as input value. This confirms the presence of an air layer under the impacting droplet.

\section*{Acknowledgments}
We would like to acknowledge the support from the foundation for Fundamental Research of Matter (FOM). We thank Mark-Jan van der Meulen for help with the MicroDrop dispenser. James Seddon is thanked for AFM measurements of the glass impact plate. We thank C. Clanet, H. Lhuissier, D. van der Meer, A. Prosperetti, J.H. Snoeijer, and P. Tsai for stimulating discussions.


\end{document}